\documentclass[dvips]{article}
\usepackage{epsfig}
\usepackage{lineno}

\title{
Anisotropy and chemical composition 
of ultra-high energy cosmic rays
using arrival directions
measured by the \\ Pierre Auger Observatory}

\date{}
\begin{document}


\maketitle

\par\noindent
{\bf The Pierre Auger Collaboration} \\
P.~Abreu$^{74}$, 
M.~Aglietta$^{57}$, 
E.J.~Ahn$^{93}$, 
I.F.M.~Albuquerque$^{19}$, 
D.~Allard$^{33}$, 
I.~Allekotte$^{1}$, 
J.~Allen$^{96}$, 
P.~Allison$^{98}$, 
J.~Alvarez Castillo$^{67}$, 
J.~Alvarez-Mu\~{n}iz$^{84}$
% ...

\par\noindent
$^{1}$ Centro At\'{o}mico Bariloche and Instituto Balseiro (CNEA-
UNCuyo-CONICET), San Carlos de Bariloche, Argentina \\
% ...
$^{19}$ Universidade de S\~{a}o Paulo, Instituto de F\'{\i}sica, S\~{a}o 
Paulo, SP, Brazil \\
% ...
$^{33}$ Laboratoire AstroParticule et Cosmologie (APC), 
Universit\'{e} Paris 7, CNRS-IN2P3, Paris, France \\
% ...
$^{57}$ Istituto di Fisica dello Spazio Interplanetario (INAF),
 Universit\`{a} di Torino and Sezione INFN, Torino, Italy \\
% ...
$^{67}$ Universidad Nacional Autonoma de Mexico, Mexico, D.F., 
Mexico \\
% ...
$^{74}$ LIP and Instituto Superior T\'{e}cnico, Lisboa, Portugal \\
% ...
$^{84}$ Universidad de Santiago de Compostela, Spain \\
% ...
$^{93}$ Fermilab, Batavia, IL, USA \\
% ...
$^{96}$ New York University, New York, NY, USA \\
% ...
$^{98}$ Ohio State University, Columbus, OH, USA \\
% ...
\par\noindent
($\dagger$) Deceased \\
(a) at Konan University, Kobe, Japan \\

\begin{abstract}
The Pierre Auger Collaboration has reported evidence for anisotropy in the
distribution of arrival directions of the cosmic rays with energies
$E>E_{th}=5.5\times 10^{19}$~eV. These show a correlation with the distribution of
nearby extragalactic objects, including an apparent excess around the
direction of Centaurus~A.  If the
particles responsible for these excesses at $E>E_{th}$ are heavy
nuclei with charge $Z$, the proton component of the sources should lead to
 excesses in the same regions at energies $E/Z$. We here
report the lack of anisotropies in these directions at energies
above $E_{th}/Z$ (for illustrative values of $Z=6,\ 13,\ 26$). 
If the anisotropies above $E_{th}$ are due to nuclei with charge $Z$, and under
reasonable assumptions about the acceleration process, these observations
imply stringent constraints on the allowed proton fraction at the
lower energies.

\end{abstract}

\vfill\eject

\section{Introduction}

Anisotropy and  composition,
together with the study of the features in the energy spectrum, are  
the fundamental tools available to decipher the origin and nature of the
ultra-high energy cosmic rays (UHECRs).
The suppression of the flux observed above 40~EeV
\cite{augerspec,hiresspec}
 suggests that the energy of the UHECRs is
attenuated by interactions with the cosmic microwave background and infrared 
photons  on their journey from their
extragalactic sources, either by photopion interactions in the case of
protons or by photodisintegration in the case of nuclei  \cite{g,zk}. 
This would imply that at the highest energies cosmic rays
can only arrive from nearby sources, within the so-called GZK horizon
(which is e.g.\ $\sim 200$~Mpc for protons above 60~EeV \cite{ha06,ol08}). 
This is supported by the correlation  reported by the Pierre Auger
Collaboration \cite{science,astropart,agnnew} between the
arrival directions of cosmic rays with energies above 55~EeV and the
distribution of nearby extragalactic objects. The correlation with nearby
active galactic nuclei (AGN) in the V\'eron-Cetty
and V\'eron (VCV) catalog \cite{vc}  was originally found with data
collected up to May 2006, and was most significant  for
the AGN within 75~Mpc and for angular separations between the AGN and cosmic
ray arrival directions  smaller than 3.1$^\circ$. A test with
subsequent data rejected the null hypothesis of isotropy with 99\% confidence
\cite{science,astropart}. A more recent analysis \cite{agnnew} has found 
that the fraction of events above 55~EeV correlating with these AGN is
($38^{+7}_{-6}$)\%, smaller than obtained initially but still well above the
isotropic expectation of 21\%. Note that these AGN
may well be acting just as tracers of the actual UHECR sources, and
indeed it is interesting that alternative studies with other populations
(X-ray AGN from the SWIFT catalog or galaxies from the 2MASS
catalog) also indicate some degree of correlation within a few
degrees with those objects \cite{agnnew} (see also
\cite{kashti,george,ghisellini,nagar,takami,tinyakov}). 
The final identification of the UHECR  sources will require much additional
data.

Another
interesting cosmic ray excess was found in the direction towards Cen~A, at  
 equatorial coordinates $(\alpha,\delta)=(201.4^\circ,-43.0^\circ )$.
Already in \cite{science,astropart} it was pointed out that two out of the 27
highest energy events observed before August 2007 by the Pierre Auger
Observatory arrived within less than $3^\circ$ of Cen~A, with several
more events 
lying in the vicinity of its radio lobes. More recently, with data up
to the end of 2009 \cite{agnnew} and considering the events above 55~EeV, 
the most significant excess around Cen~A
was identified for an 18$^\circ$
window, in which 13 events were observed while only 3.2 were expected.
Whether this excess, if confirmed with further data, is due to Cen~A,
which is one of the nearest AGN (being at less than 4~Mpc
distance), or due to one or several sources farther away, e.g.\ in the
Centaurus cluster lying in a similar direction but at $\sim 45$~Mpc,
is something that remains to be determined. It should be mentioned
that the HiRes air shower experiment has not found
indications of an excess correlation with nearby AGN \cite{hiresagn},
although the associated statistics are smaller and there are systematic
differences in the energy calibrations between the two experiments. 
Also, contrary to Auger, the HiRes experiment
 looks to the northern hemisphere, and in particular this makes it
 blind to the Cen~A region of the sky.

The Pierre Auger Observatory has recently measured the average depth of the
maximum of shower development $X_{max}$ and its fluctuations \cite{augercomp}.  The
logarithmic slope of the average shower maximum vs. energy becomes smaller
above $\sim 2$~EeV, indicating a change in the shower properties. Also the
fluctuations in $X_{max}$ become suppressed above this energy. 
  An inference of the chemical composition of the primary cosmic rays can be done
  via comparison with Monte Carlo simulations of air showers. If these models
  are taken at face value, they indicate a gradual increase in
  the average mass as a function of energy\footnote{
We note that the HiRes experiment measures a depth of shower maximum
consistent with proton-only Monte Carlo air shower simulations 
 all the way from 1~EeV up to $\sim 40$~EeV \cite{hirescomp}.}. 
 Alternatively, this behavior
  could be ascribed to changes in the hadronic interactions (cross sections,
  inelasticities or multiplicities)  not considered in the available 
models. We note that the models make extrapolations to energies well beyond
those tested at accelerators.
One should also keep in mind that, due to the limited statistics of the events
observed with fluorescence telescopes, there is no measurement of the mean
$X_{max}$ and the corresponding fluctuations available for $E > 55$~EeV. 
  It
  is clear that performing alternative studies to try to improve our
  understanding of the UHECR composition is important.

In this work we perform searches  for anisotropies in the same directions where
excesses were observed above $E_{th}=55$~EeV, but using lower energy
thresholds (we consider the illustrative values for the threshold $E_{th}/Z$,
with $Z=6,\ 13,\ 26$).

We first focus on the analysis of the region  around
Cen~A, for which the most significant excess 
was found above $55$~EeV for an angular
window of  $18^\circ$ radius.
We note that the location of the excess, the size of the angular window and the
selected energy threshold are a posteriori, therefore
 new independent data would be required
to assess the  significance of the excess at high energies.
However, we are already able to
report the results of our search for anisotropies 
in the same region for the lower  energy thresholds considered.

We  also perform a similar search but  looking for possible excesses in
windows of 3.1$^\circ$ around the VCV AGN  within 75~Mpc, considering
only the data after May 2006 so as to exclude those used to 
fix these parameters. 
We note that even above 55~EeV the cosmic ray deflections
in the galactic magnetic field are likely larger than  a few degrees
(especially if cosmic rays happen to be heavy nuclei).  However, the VCV
correlation does not imply that the objects in this catalog are the sources,
nor that the typical deflections are 
smaller than the optimal correlation angle. Active galaxies in
the VCV catalog trace 
the nearby large scale matter distribution, and that 
includes all types of candidate astrophysical sources, not only AGN
and their subclasses. Deflections of cosmic ray trajectories could be larger,
and still manifest an anisotropy  through a correlation of a fraction of them
within a few degrees of the VCV objects. 

We then explore the possibility that the anisotropies at the highest energies
might be due to heavy nuclei.   Using our observations and following an idea
proposed by Lemoine and Waxman  \cite{le09},   
which exploits the fact that a high energy anisotropy due to nuclei of charge
$Z$  should lead to an anisotropy in the same region of the sky 
at energies $Z$ times smaller due to the
protons from the same sources,  we are then able to
constrain the allowed proton fraction at the source under different
assumptions on the value of the nuclear charges responsible for the high
energy excess.

\section{The Observatory and the dataset:}

The Pierre Auger Observatory is located near the town of Malarg\"ue,
Argentina, at a latitude of 35.25$^\circ$~S. It is a hybrid detector,
consisting of 24 fluorescence telescopes and a surface array of 1600
water Cherenkov detectors covering $\sim 3000$~km$^2$ (see
\cite{auger,augerfd} for further details).  

The data considered in the present
work consists of the cosmic ray events  with zenith angles $\theta<60^\circ$
detected by the surface array
(which has an almost 100\% duty cycle and hence collected the largest data set)
since 1 January 2004 up to 31 December 2009. 
The array has been growing in size until the
completion of the baseline design 
in mid  2008. In order to have an accurate estimate
of the exposure and hence of the expected background in the different
regions of the sky  we have removed
periods in which the data acquisition was unstable (the 
resulting livetime being 87\% \cite{auger}) 
and applied  a quality cut that requires that for any event  
the six detectors surrounding
the detector having the largest signal  be active at the
time the event is recorded.
Keeping track of the number of active
detector configurations able to trigger such events 
 at any time allows us  to take into account the detector
growth and dead times in the evaluation of the exposure. 
 The isotropic expectation in an  angular window  $\Delta\Omega$ can be obtained
 as   $N_{iso}=x\ N_{tot}$, where $x$ is the fraction  
of the exposure within the solid angle $\Delta\Omega$ and $N_{tot}$ is the
total number of events.

 The trigger efficency is 100\%  for $E>3$~EeV, but at 
lower energies (we consider here events down
to $E=55$~EeV$/26 \simeq 2.1$~EeV) the trigger efficiency becomes
smaller than unity and is zenith angle dependent. Hence, to obtain the
isotropic expectations for the lower energy threshold considered
 we use a fit to the zenith angle distribution
of the events, rather than the  ideal exposure  expectation d$N\propto
\sin\theta\cos\theta {\rm d}\theta$.  We note that the detection 
efficiency below 3~EeV may also depend on the composition of the
cosmic rays, being actually smaller for lighter nuclei. 
This could slightly affect the predictions for the  
expected localized proton excesses 
 for the lowest energy threshold
considered, $E>2.1$~EeV. We estimate that in this case 
the predictions are affected
by no more than 2\% by the possible differences in exposure (using the values
in ref.~\cite{auger}), and hence these
effects can be safely neglected.

\section{Results}
\subsection{The Centaurus A excess}

\begin{figure}[htpb]
\centerline{\includegraphics[angle=-90,width=0.6\linewidth]{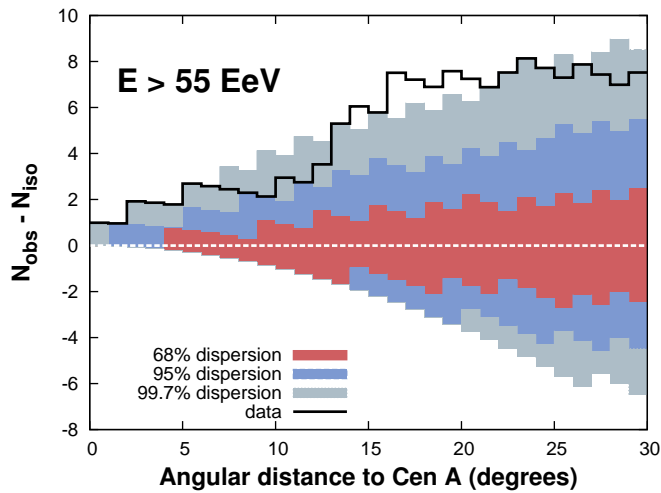}}
\caption{Cumulative number of events with $E\geq 55$ EeV 
(subtracting the average  isotropic expectations) as a function
of angular distance from the direction of Cen A. The bands correspond to the 68\%, 95\% and 99.7\% dispersion 
expected for an isotropic flux.}
\label{figure-CenA55}
\end{figure}

We first consider the excess observed in the Cen~A region
for energies above the threshold $E_{th}=55$~EeV.  
The cumulative number of events as a function
of the angular distance from the direction of Cen A is plotted in
Fig.~\ref{figure-CenA55} (to make the plot more readable we
display the difference with respect to the average 
isotropic expectations). 

\begin{figure}[htpb]
\centerline{\includegraphics[angle=-90,width=0.5\linewidth]{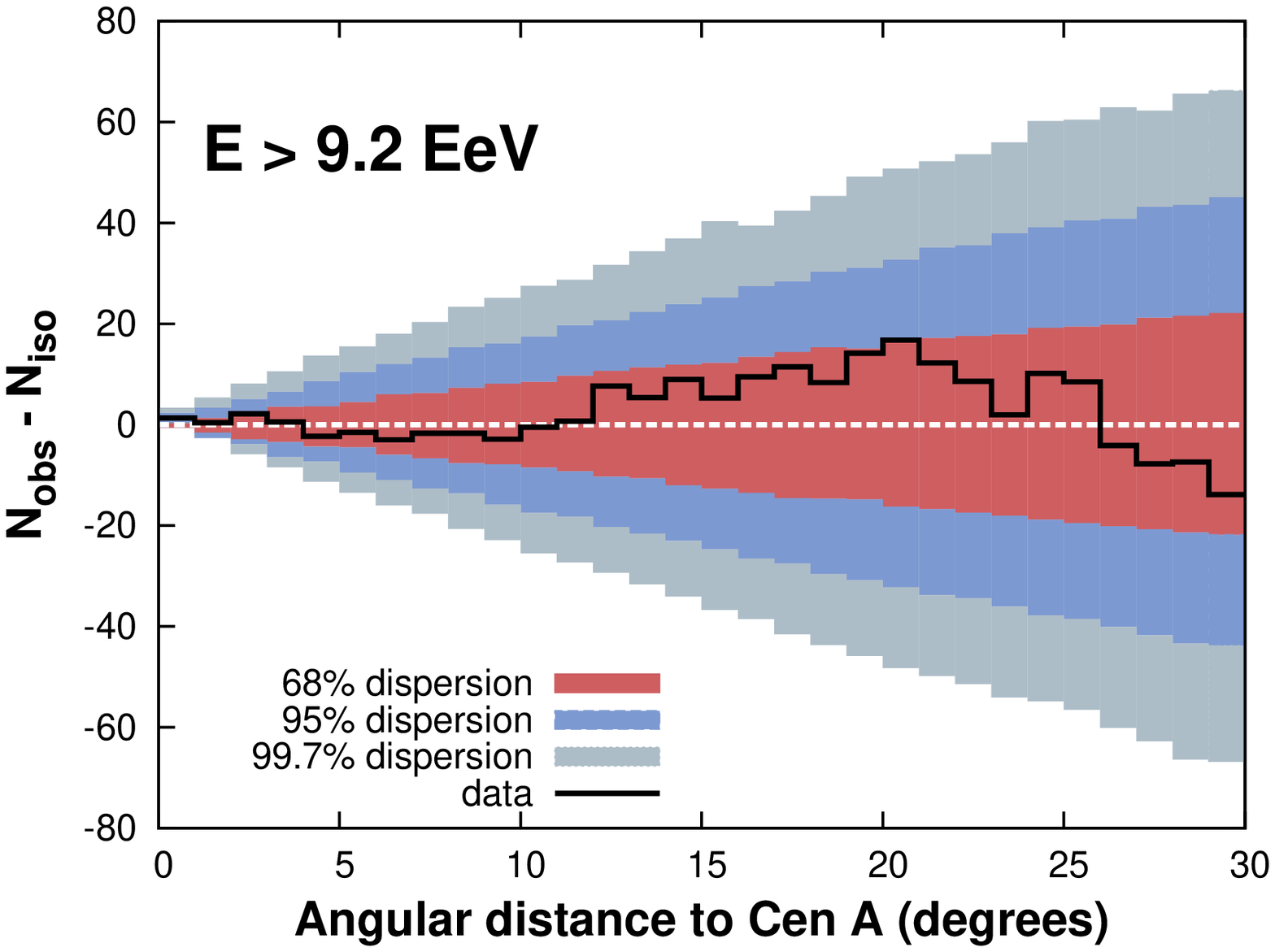}}
\medskip
\centerline{\includegraphics[angle=-90,width=0.5\linewidth]{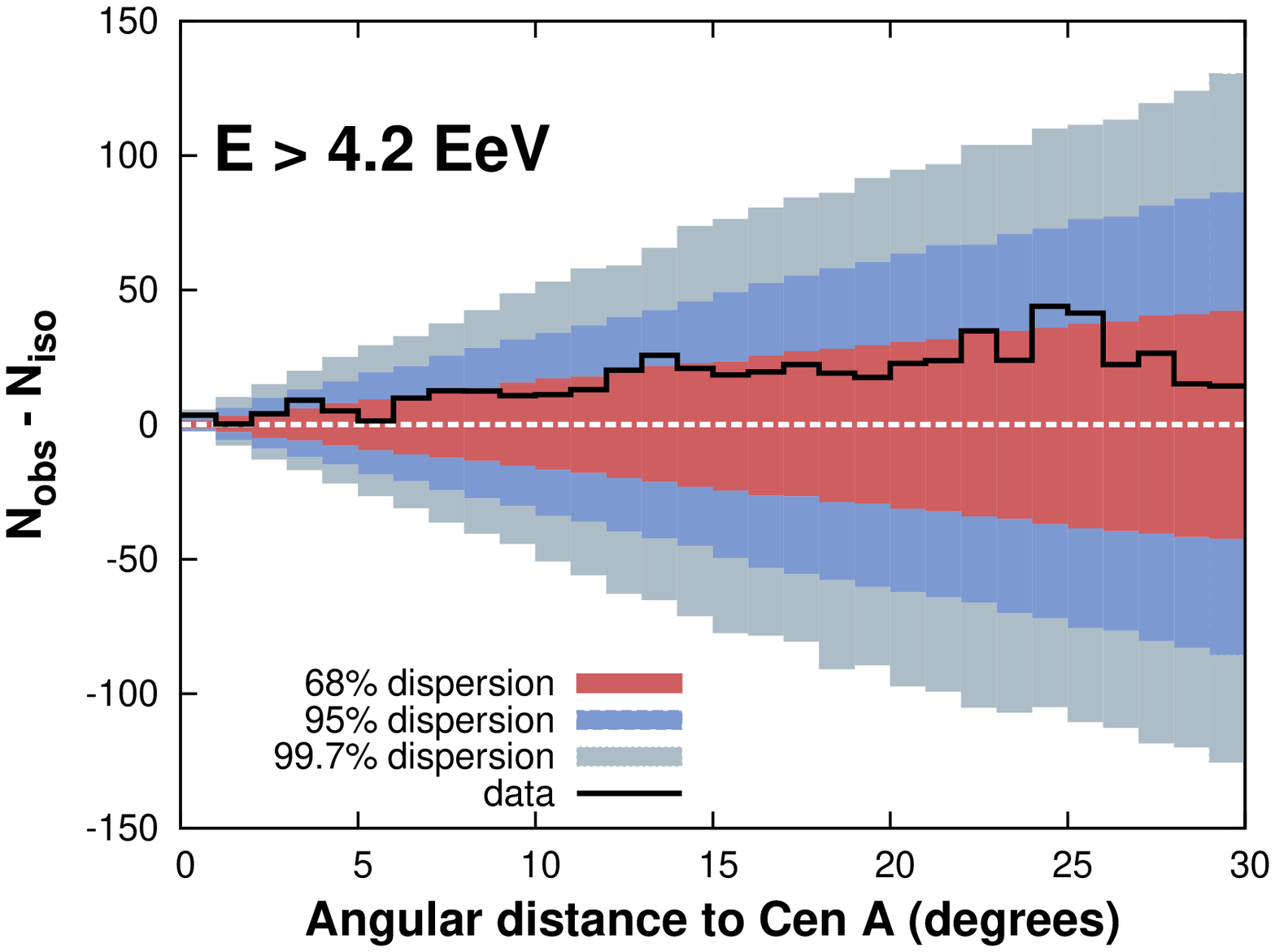}
\includegraphics[angle=-90,width=0.5\linewidth]{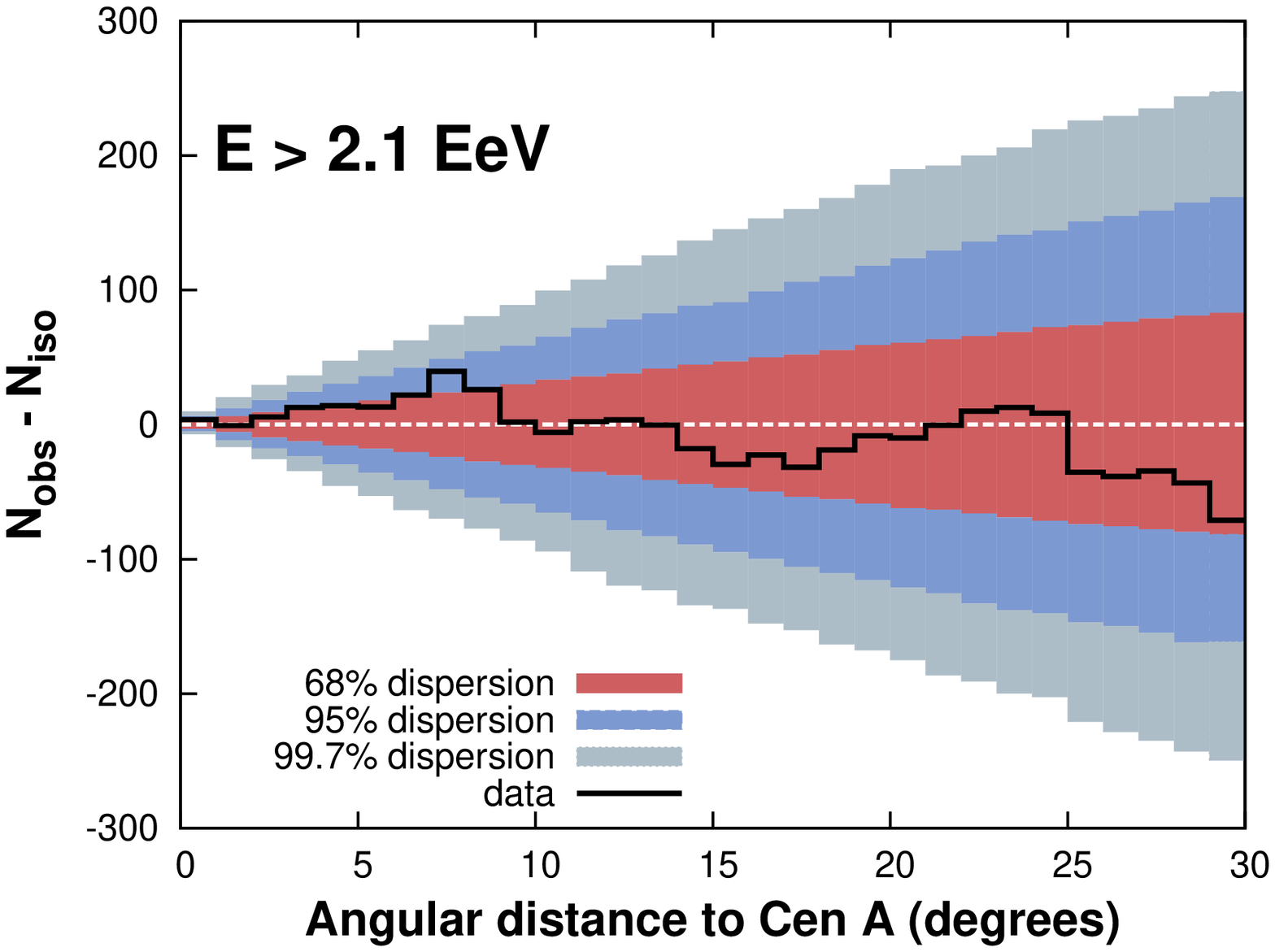}}
\caption{Similar to Fig.~\ref{figure-CenA55},
 events with $E\geq 55~{\rm EeV}/Z$ for $Z=6$ (top), 13 (bottom left) and 26
 (bottom right).
}
\label{figure-CenAZ}
\end{figure}

In Fig.~\ref{figure-CenAZ} we plot the cumulative number of events,
subtracting the isotropic expectations,  as a function
of the angular distance from the direction of Cen A for lower energy
thresholds, considering energies above
$E_{th}/Z$ in the  cases $Z=6$, 13 and 26. The observed distributions are
consistent with the isotropic expectations (shaded regions), showing  no
significant excesses in any of the angular windows considered.

As reported in \cite{agnnew}, the most significant
excess for a top-hat window  around Cen~A 
was obtained for a radius $\gamma=18^\circ$ and we will hence focus on this
region. 
For this  energy range, 
the total number of events\footnote{ Different from ref.~\cite{agnnew}, 
where 13 out of 69 events were reported to correlate 
within $18^\circ$ of Cen~A, the stricter event selection applied in this work 
in order to get an accurate estimate of the exposure at low energies yields 10 
correlations out of 60 events, well within the statistical uncertainties of 
the previous result.}
  is  $N_{tot}=60$,
with $N_{obs} = 10$ of these being in an $18^\circ$ angular window around Cen~A.
 If we adopt the expression for the ideal exposure of the detector, 
the fraction of isotropic sky in
this $18^\circ$ region is $x \simeq 0.0466$. Normalizing to the counts outside
the ‘source’ region, the expected background in this region is $N_{bkg} =
(N_{tot}-N_{obs})x/(1-x) = 2.44$ counts.

In table~\ref{table-CenA} we report the observed number of events  with 
$E > 55$ EeV/Z   (total and in an angular 
window of $18^\circ$ around Cen~A), as well as the expected isotropic
background.  No significant excess is found for any 
of the lower energy thresholds considered.

\begin{table}[htpb]
\centerline{\begin{tabular}{|c|c||c|c|c|c|}
\hline
$Z$ & $E_{min}$ [EeV] & $N_{tot}$ & $N_{obs}$ & $N_{bkg}$  \\
\hline \hline 
6 & 9.2 &  4455   &  219  &$ 207\pm 14$   \\
\hline
13 & 4.2 & 16640  & 797 & $774\pm 28$   \\
\hline
26 & 2.1 & 63600 & 2887 & $2920\pm 54$  \\
\hline
\end{tabular}}
\caption{Total number of events, $N_{tot}$, and those observed in an angular
window of $18^\circ$ around Cen~A, $N_{obs}$, 
as well as the expected background $N_{bkg}$. 
Results are given  for different
energy thresholds, corresponding to $E_{min}=E_{th}/Z$ for the indicated values
of $Z$ and $E_{th}=55$~EeV. }
\label{table-CenA}
\end{table}

\subsection{The VCV AGN}

We now search for possible overdensities of cosmic rays 
with arrival directions within $3.1^\circ$ of objects with
redshift $z\le 0.018$ ($\sim 75$ Mpc) in the  
VCV catalog. We use for this study only data collected after May 2006, 
subsequent to data used to specify the parameters that optimized the VCV
correlation in that period. 

In this case, one has that for $E>E_{th}=55$~ EeV there are $N_{tot}=49$ events, of which 
$N_{obs}=20$ are within $3.1^\circ$ of the nearby AGN. 
On the other hand, the probability that
isotropic cosmic rays correlate by chance with those 
objects is $x\simeq 0.212$ and hence
$N_{bkg}=(N_{tot}-N_{obs})x/(1-x)= 7.88$. 

In table~\ref{Table-VCV} we show the observed number of events with $E >
55~{\rm EeV}/Z$ (total and those within $3.1^\circ$ of an object with $z\le
0.018$ in the VCV catalog),  as well as the expected background.
It is apparent that no significant excess is found for
 any of the lower energy thresholds considered.

\begin{table}[htpb]
\centerline{\begin{tabular}{|c|c||c|c|c|}
\hline
$Z$ & $E_{min}$ [EeV] & $N_{tot}$ & $N_{obs}$ & $N_{bkg}$  \\
\hline \hline 
6 & 9.2 &  3626   &  763  & $770\pm 28$    \\
\hline
13 & 4.2 & 13482  & 2852 & $2860\pm 54$   \\
\hline
26 & 2.1 & 51641 & 10881 & $10966\pm 105$  \\
\hline
\end{tabular}}
\caption{Total number of events, $N_{tot}$, and those observed 
within $3.1^\circ$ from 
objects with $z\le 0.018$ in the VCV catalog, $N_{obs}$, as well as the expected
isotropic background  $N_{bkg}$.
Results are given for different 
energy thresholds, corresponding to $E_{min}=E_{th}/Z$ for the indicated values
of $Z$ and $E_{th}=55$~EeV. }
\label{Table-VCV}
\end{table}

\section{Constraints on the source composition}

As a by-product of the observations described above, and under reasonable
assumptions on the cosmic ray acceleration and propagation,  it is possible to set
some constraints on the composition  of the cosmic rays responsible
for localized overdensities observed  above $E_{th}$. In order to do this, we
elaborate on an idea proposed by  Lemoine and Waxman \cite{le09}, who related
the high energy excess, 
under the assumption that it is due to heavy nuclei of charge $Z$, with the
expected excess at energies above $E_{th}/Z$ due to the protons from the same
sources. Note that, in the absence of energy losses and scattering effects, 
 protons with energies $E/Z$ would follow the same trajectories as 
nuclei of charge $Z$ and energy $E$  coming from the same source, and hence
they  should arrive within the same
angular windows. Moreover,  even if at lower energies the isotropic background 
can be enhanced by the contribution from sources beyond the GZK horizon,
the gain in statistics obtained  can make the search  sensitive to relatively
small low energy anisotropies.

The main underlying hypothesis is that the cosmic ray acceleration depends
just on the particle rigidities, i.e.\ on $E/Z$. It is therefore 
natural to assume that
at the sources the spectra of the different charge components 
scale as
\begin{equation}
\frac{{\rm d}n_Z}{{\rm d}E}=k_Z \Phi( E/Z),
\label{spectra}
\end{equation}
with $k_Z$ being constant factors.
The function $\Phi$ may display a high energy cutoff resulting from
the maximum rigidities attainable by the acceleration process. If, in this
scenario, the maximum proton energies were below $E_{th}$, 
the higher energy cosmic
rays from the source could be dominated by a heavy component.

If $N(>E)$ is the number of events with energies above the threshold $E$ which  
come  within a certain solid
angle around a source and if the acceleration process at the source depends only on
rigidity, then the number 
of nuclei of charge $Z$ above $E_{th}$ and those of
protons above $E_{th}/Z$ are related by
\begin{equation}
N_p(>E_{th}/Z)=\frac{k_p}{Zk_Z} 
N_Z(>E_{th}).
\label{npz}
\end{equation}
This relation does not take energy losses into account
 (included as a parameter $\alpha$ in
ref.~\cite{le09}). Ignoring them leads to more  conservative bounds on the
ratio $k_p/k_Z$,
because energy losses are larger for nuclei of charge $Z$ and energy $E$ than
for protons of energy $E/Z$. Moreover, the nucleons emitted in the
photodisintegration processes can also add to the expected proton anisotropies
at low energies.

The number of events produced by the 
source(s) responsible for the localized excess observed  can 
be estimated as $N=N_{obs}-N_{bkg}$ in terms of the number of events
observed in the window considered and the expected background, which are
displayed in the tables.
Taking into account the Poisson fluctuations in the low and high energy
signals, as well as in the background estimates, we obtain 95\% CL
upperbounds on the 
quantity $R_Z\equiv N(>E_{th}/Z)/N(>E_{th})$ using the
profile likelihood method (see e.g.\ \cite{rolke}).
 In the case of Cen~A these
bounds are $R_Z<12.9$, 17.3 and 9.1 for $Z=26$, 13 and 6 respectively, while
for the case of VCV the bounds are $R_Z<14.7$, 12.4 and 6.0. We note that
considering a 99\% confidence level, the bounds become $R_Z< 23.8$, 31.1 and
16.3  for Cen~A, and $R_Z< 28.9$, 23.7 and 11.4 for VCV (for $Z=26$,
  13 and 6 respectively), being then typically a factor of two weaker. 

If the excess at high energies is indeed 
dominated by the heavier nuclear component
of charge $Z$, i.e.\ $N(>E_{th})\simeq N_Z(>E_{th})$, we obtain that $R_Z>
N_p(>E_{th}/Z)/N_Z(>E_{th})+1=k_p/(Zk_Z)+1$ (where we used that
$N_Z(>E_{th}/Z)/N_Z(>E_{th})>1)$.  In this way, conservative bounds
$k_p/k_Z<Z(R_Z-1)$ can be obtained.

One may translate these limits on the relative spectrum normalizations
 into bounds on the actual low energy 
 abundance ratios between the proton and heavy elements at the source.
In particular, in the case that one assumes that below a certain rigidity 
the spectrum has a power law behavior, i.e. $\Phi \propto (E/Z)^{-s}$ for 
$E/Z<E_1$, as expected in
scenarios of diffusive shock acceleration, at energies below $E_1$ all the
relative abundances of the different elements present will be
independent of the energy. 
In this case,  one can relate the low energy relative
fractions $f_i$ of the different elements at the source with  the
normalization factors $k_i$ in eq.~(\ref{spectra}). 
Comparing the differential spectra
 for protons and for the charge $Z$ at energies below $E_1$ one gets
\begin{equation}
\frac{k_p}{k_Z}=\frac{f_p}{f_Z}Z^{s}.
\end{equation}
Note that we are not making any assumption about the spectral shape above
the threshold energies. Also,  for energies above $E_1$ the values of
$f_p/f_Z$ will depend on the spectral shape details.

\begin{figure}[tpb]
\centerline{\includegraphics[width=0.7\linewidth]{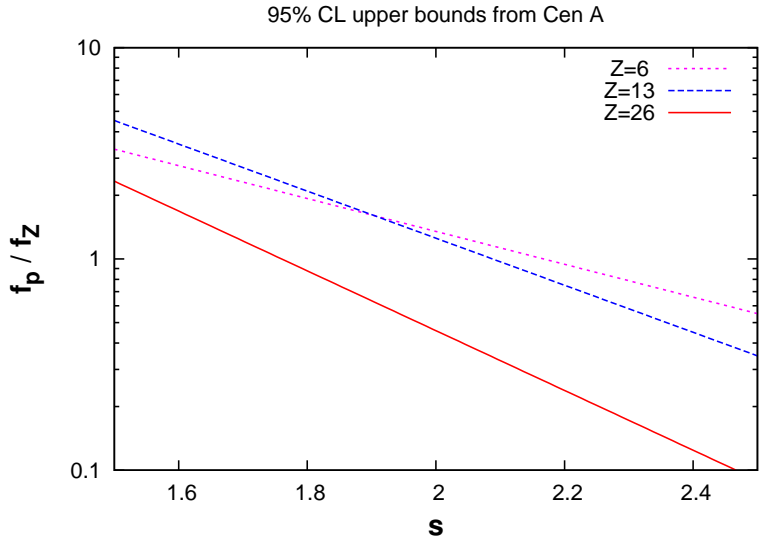}}
\caption{Upper bounds at 95\%CL 
on the allowed proton to heavy fractions in the source as a
  function of the assumed low energy spectral index $s$. 
 The different lines are for charges $Z=6$, 13 and 26, as
indicated.} 
\label{fpCenA}
\end{figure}

\begin{figure}[htpb]
\centerline{\includegraphics[width=0.7\linewidth]{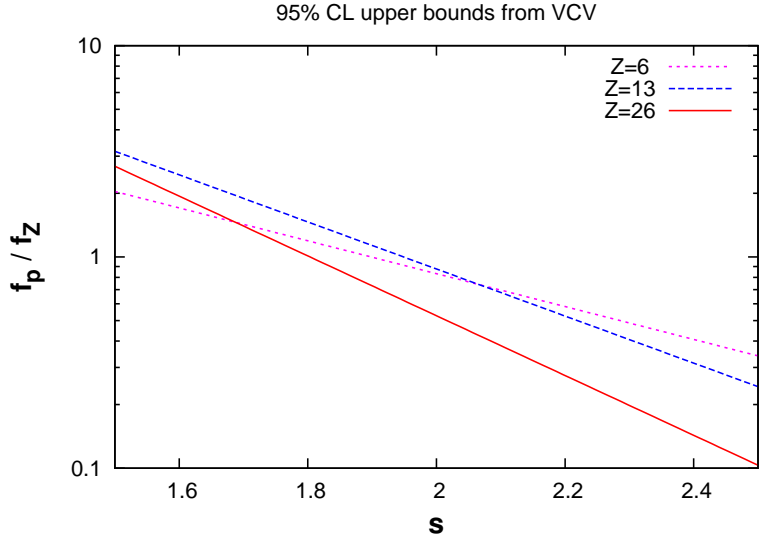} 
}
\caption{Upper bounds at 95\% CL
on the allowed proton to heavy fractions in the source as a
  function of the assumed low energy spectral index $s$ implied by the absence 
of an excess of arrival directions of cosmic rays with energies above 55~
EeV/$Z$  
within $3.1^\circ$ of objects with $z\le 0.018$ in the VCV catalog.  The
different lines are for charges $Z=6,$ 13 and 26, as indicated.} 
\label{fpVCV}
\end{figure}

The resulting bounds for the low energy relative abundances are displayed in 
Figs. ~\ref{fpCenA} and \ref{fpVCV} as a function of the low energy spectral
index, for values   $1.5<s<2.5$. The regions above the respective lines
are excluded at 95\% CL.

In the case in which energy losses can be neglected, such as if the source of
the excess events is the nearby Cen~A galaxy,
 it is appropriate to consider energy bins and relate, through an expression
 analogous to that in eq.~(\ref{npz}),
 the events in the bin $[E_{th},2E_{th}]$ to
 those at $Z$ times lower energies (where we adopted for definiteness
 a bin width corresponding to a factor two in energy)\footnote{If energy losses
 were relevant,
 the observed energies of the high energy events might correspond to a wider
 span of energies at the source, and the corresponding low energy protons may
 then
 span a range of energies wider than $[E_{th}/Z,2E_{th}/Z]$, making this
 analysis no longer valid, while that based on the integral energy bins
 above a threshold would still provide conservative bounds.}.
 The ratio between the events observed and those expected for the $18^\circ$
 window around Cen~A are $N_{obs}/N_{bkg}=152/153.5$, 543/533.4 and
 2090/2147.9 for 
 $Z=6$, 13 and 26 respectively. This leads to bounds on the proton
 fractions similar to those in Fig.~\ref{fpCenA}
 but about a factor two stronger.

An important point is that the statistical significance of the constraints in
Fig.~\ref{fpCenA} is a posteriori, since 
the identification of the region  around Cen~A,
 its angular size  and
the energy threshold were tuned to maximize this excess. Therefore it 
would be necessary to look to this same region using the 
same energy threshold with an independent dataset of  comparable size 
so as to obtain an unbiased estimate of the strength  of the source (or
sources) producing the excess. 
We note however that varying the energy threshold to 50 or 60~EeV leads to
qualitatively similar results. Also the angular size adopted for 
the window is
not very crucial. For instance, if we consider a
 $10^\circ$ window 
 instead of the $18^\circ$ one, the main effect on the bounds comes 
from the modification of
 the expected background in the low energy bin. This would
relax the bounds on $f_p/f_{Fe}$ by a factor of about two in this case.
It is interesting to point out  that the sensitivity of the a
posteriori bounds from Cen~A turns out to be
 comparable to that achieved with the analysis
of the VCV correlations.

\section{Discussion and conclusions}

We have searched for overdensities at energies
$E_{th}/Z$ in the regions where anisotropies were reported previously above
 $E_{th}=55$~EeV, i.e.\ both in the direction towards Cen~A and in
3.1$^\circ$ windows around nearby AGN from the VCV catalog. 
Considering representative values of $Z=6$, 13 and 26, where $Z$
is the assumed charge of the cosmic rays responsible for the high
energy anisotropies, we have found no indications of overdensities in
any of the lower energy bins.
In scenarios where the acceleration process is only dependent on rigidity,   
the absence of significant anisotropies at energies $E>E_{th}/Z$
implies  that  an upper bound can be set on the
low-energy relative proton abundance at 
the sources. This bound is given by $f_p/f_Z \leq (0.5$ to $2)Z^{2-s}$ at the
95\% CL level, depending on the adopted value of $Z$ (see 
lines in Figs.~\ref{fpCenA} and \ref{fpVCV}).  
Note that the
 constraints become weaker if the source
 spectrum is very hard ($s\simeq 1.5$). Given the comparable
  bounds obtained for different values of $Z$, similar limits will result in
  the case in which the high energy anisotropy is dominated by nuclei
  belonging to a given mass group with similar values of $Z$.

On the other hand, estimates of the expected low energy relative
abundances point towards values
 above these bounds. For instance, the ATIC-2
experiment \cite{pa06} has measured that at 100~TeV (the highest energies for
which
detailed composition measurements are available) one has $f_p\simeq
f_{He}\simeq 2f_{CNO}\simeq 2f_{Ne-Si}\simeq 2f_{Z>17}\simeq 4
f_{Fe}$. 
Moreover, for these low energies (for which cosmic rays are believed to be
of galactic origin) one would expect that the measured relative 
fraction of protons versus heavy nuclei for a given particle energy 
is actually smaller than the original fraction at the sources, due to the longer
confinement time  in the Galaxy of the heavier species. For instance,
 the measured $p$ to Fe fraction would be $26^{1/3}\simeq
3$ times smaller than the value at the source if the turbulent component of the
galactic magnetic field has a Kolmogorov spectrum, so that the
diffusion coefficient scales as $D\propto (E/Z)^{1/3}$. One has to keep in
mind that these estimates based on lower energy 
galactic cosmic ray sources do not
necessarily apply to the extragalactic sources which are most likely
responsible for the highest energy events, but one may consider that 
they provide a useful indication of the plausible expected values. 

Hence, we conclude that 
a heavy composition for the excesses observed at high energies
appears to be  in
conflict with rigidity-dependent acceleration scenarios having at
low energies a proton component more abundant  than heavier species, 
as quantified in
Fig.~\ref{fpVCV}.
 How these conclusions are modified in the 
presence of strong structured magnetic fields and taking into account the 
relevant energy losses remains to be seen.
We note that the present 
analysis based on the lack of anisotropies at lower energies provides
information on the cosmic ray composition which is  
 independent of  $X_{max}$ measurements, but depends instead 
on assumptions related to source properties.

\section*{Acknowledgments}

The successful installation and commissioning of the Pierre Auger Observatory
would not have been possible without the strong commitment and effort
from the technical and administrative staff in Malarg\"ue.

We are very grateful to the following agencies and organizations for financial support: 
Comisi\'on Nacional de Energ\'{\i}a At\'omica, 
Fundaci\'on Antorchas,
Gobierno De La Provincia de Mendoza, 
Municipalidad de Malarg\"ue,
NDM Holdings and Valle Las Le\~nas, in gratitude for their continuing
cooperation over land access, Argentina; 
the Australian Research Council;
Conselho Nacional de Desenvolvimento Cient\'{\i}fico e Tecnol\'ogico (CNPq),
Financiadora de Estudos e Projetos (FINEP),
Funda\c{c}\~ao de Amparo \`a Pesquisa do Estado de Rio de Janeiro (FAPERJ),
Funda\c{c}\~ao de Amparo \`a Pesquisa do Estado de S\~ao Paulo (FAPESP),
Minist\'erio de Ci\^{e}ncia e Tecnologia (MCT), Brazil;
AVCR, AV0Z10100502 and AV0Z10100522,
GAAV KJB300100801 and KJB100100904,
MSMT-CR LA08016, LC527, 1M06002, and MSM0021620859, Czech Republic;
Centre de Calcul IN2P3/CNRS, 
Centre National de la Recherche Scientifique (CNRS),
Conseil R\'egional Ile-de-France,
D\'epartement  Physique Nucl\'eaire et Corpusculaire (PNC-IN2P3/CNRS),
D\'epartement Sciences de l'Univers (SDU-INSU/CNRS), France;
Bundesministerium f\"ur Bildung und Forschung (BMBF),
Deutsche Forschungsgemeinschaft (DFG),
Finanzministerium Baden-W\"urttemberg,
Helmholtz-Gemeinschaft Deutscher Forschungszentren (HGF),
Ministerium f\"ur Wissenschaft und Forschung, Nordrhein-Westfalen,
Ministerium f\"ur Wissenschaft, Forschung und Kunst, Baden-W\"urttemberg, Germany; 
Istituto Nazionale di Fisica Nucleare (INFN),
Istituto Nazionale di Astrofisica (INAF),
Ministero dell'Istruzione, dell'Universit\`a e della Ricerca (MIUR), 
Gran Sasso Center for Astroparticle Physics (CFA), Italy;
Consejo Nacional de Ciencia y Tecnolog\'{\i}a (CONACYT), Mexico;
Ministerie van Onderwijs, Cultuur en Wetenschap,
Nederlandse Organisatie voor Wetenschappelijk Onderzoek (NWO),
Stichting voor Fundamenteel Onderzoek der Materie (FOM), Netherlands;
Ministry of Science and Higher Education,
Grant Nos. 1 P03 D 014 30 and N N202 207238, Poland;
Funda\c{c}\~ao para a Ci\^{e}ncia e a Tecnologia, Portugal;
Ministry for Higher Education, Science, and Technology,
Slovenian Research Agency, Slovenia;
Comunidad de Madrid, 
Consejer\'{\i}a de Educaci\'on de la Comunidad de Castilla La Mancha, 
FEDER funds, 
Ministerio de Ciencia e Innovaci\'on and Consolider-Ingenio 2010 (CPAN),
Generalitat Valenciana, 
Junta de Andaluc\'{\i}a, 
Xunta de Galicia, Spain;
Science and Technology Facilities Council, United Kingdom;
Department of Energy, Contract Nos. DE-AC02-07CH11359, DE-FR02-04ER41300,
National Science Foundation, Grant No. 0969400,
The Grainger Foundation USA; 
NAFOSTED, Vietnam;
ALFA-EC / HELEN,
European Union 6th Framework Program,
Grant No. MEIF-CT-2005-025057, 
European Union 7th Framework Program, Grant No. PIEF-GA-2008-220240,
and UNESCO.


\end{document}